\documentclass[reprint,superscriptaddress,amsmath,amssymb,aps]{revtex4-2}
\usepackage{graphicx}
\usepackage{dcolumn}
\usepackage{bm}
\usepackage[colorlinks,linkcolor=blue,citecolor=blue]{hyperref}

\begin{document}

\title{Linear elasticity of polymer gels in terms of negative energy elasticity}

\author{Naoyuki~Sakumichi}
\email{sakumichi@tetrapod.t.u-tokyo.ac.jp}
\affiliation{Department of Bioengineering, Graduate School of Engineering, The University of Tokyo, 7-3-1 Hongo, Bunkyo-ku, Tokyo, Japan.}
\author{Yuki~Yoshikawa}
\affiliation{Department of Bioengineering, Graduate School of Engineering, The University of Tokyo, 7-3-1 Hongo, Bunkyo-ku, Tokyo, Japan.}
\author{Takamasa~Sakai}
\email{sakai@tetrapod.t.u-tokyo.ac.jp}
\affiliation{Department of Bioengineering, Graduate School of Engineering, The University of Tokyo, 7-3-1 Hongo, Bunkyo-ku, Tokyo, Japan.}
\date{\today}

\begin{abstract}
We recently found that the energy contribution to the linear elasticity of polymer gels in the as-prepared state can be a significant negative value; the shear modulus is not proportional to the absolute temperature [Yoshikawa Y et al., Phys. Rev. X \textbf{11}, 011045 (2021)].
Our finding challenges the conventional notion that the polymer-gel elasticity is mainly determined by the entropy contribution.
Existing molecular models of classical rubber elasticity theories, including the affine, phantom, and junction affine network models, cannot be used to estimate the structural parameters of polymer gels.
In this focus review, we summarize the experimental studies on the linear elasticity of polymer gels in the as-prepared state using tetra-arm poly(ethylene glycol) (PEG) hydrogels with a homogenous polymer network.
We also provide a unified formula for the linear elasticity of polymer gels with various network topologies and densities.
Using the unified formula, we reconcile the past experimental results that seemed to be inconsistent with each other.
Finally, we mention that there are still fundamental unresolved problems involving the linear elasticity of polymer gels.
\end{abstract}

\maketitle

\section{Introduction}

Polymer gels are widely used in food products such as yogurt, tofu, and jelly~\cite{Baziwane2003, Saha2010, Peng2015} and in biomaterials such as anti-adhesion agents, hemostatic agents, and soft contact lenses~\cite{Yeo2007, Gaharwar2014, Calo2015}. 
For these applications, it is important to control the stiffness of polymer gels.
For example, flexible polymer gels are used in artificial vitreous substitutes and food for dysphagia, and stiff polymer gels are used in hemostatic agents and artificial cartilage.
By optimizing the polymer gel stiffness for its intended use, the quality of life (QOL) can be improved in various situations.\\

Despite the importance of controlling the stiffness, it is an open question how the stiffness of a polymer gel is determined by its microscopic network structure.
The elastic behavior of polymer gels, which are usually regarded as rubber containing a large amount of solvent, has been conventionally analyzed and predicted based on models of classical rubber elasticity theories, such as the affine~\cite{Flory1953}, phantom~\cite{James1953}, and junction affine network models~\cite{Flory1977}.
However, it is difficult to verify the applicability of these microscopic models to the macroscopic properties of polymer gels because conventional polymer gels inherently have inhomogeneous network structures~\cite{Shibayama1998}.
Thus, the determination of the appropriate microscopic model describing polymer-gel elasticity remains to be achieved~\cite{Patel1992, Hild1998}.\\

\begin{figure}[t!]
\centering
\includegraphics[width=0.89\linewidth]{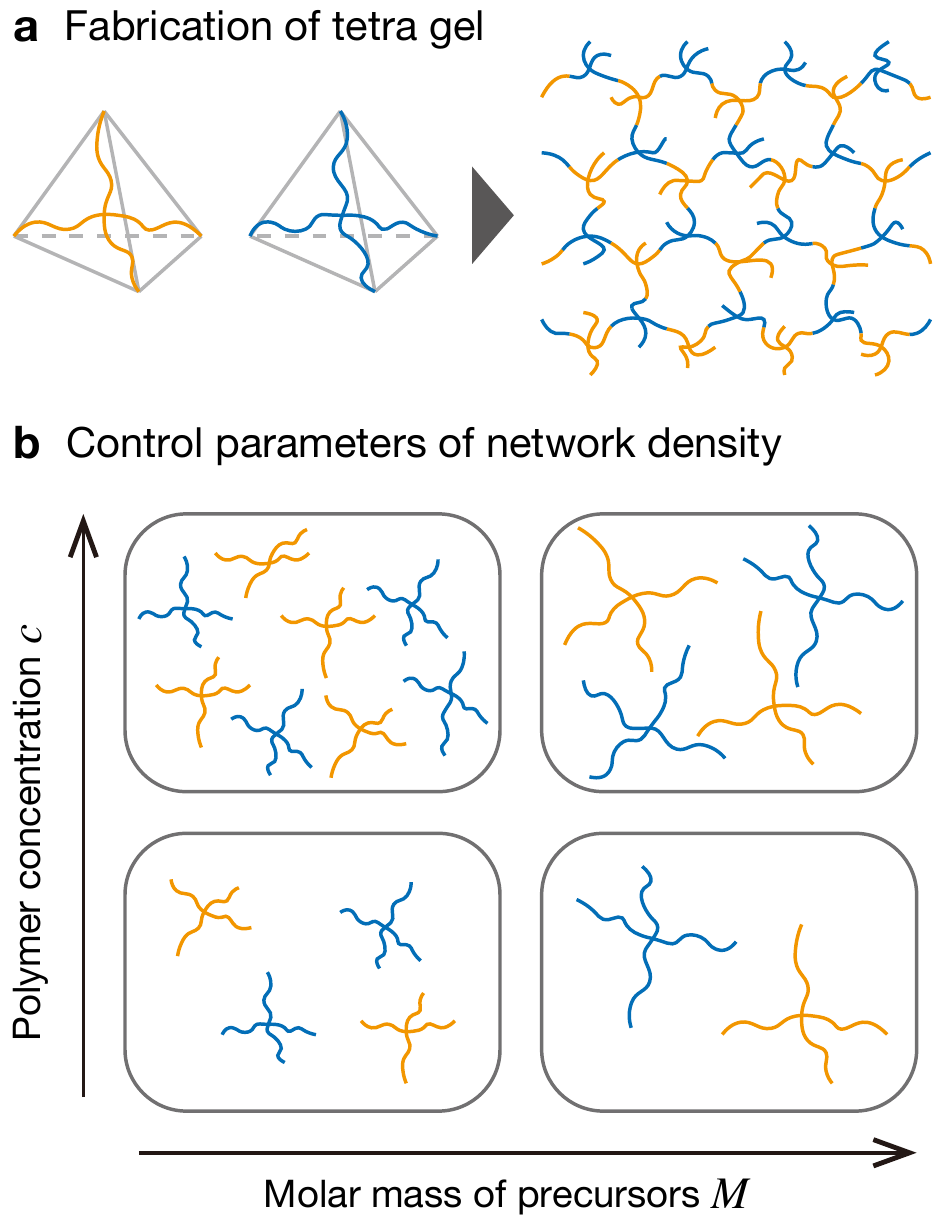}
\caption{
\textbf{a}
Tetra gel synthesized by AB-type cross-end coupling of two kinds of precursors of equal size in a water solvent.
These precursors are tetra-arm poly(ethylene glycol) (PEG) chains whose terminal functional groups (A and B) are mutually reactive.\\
\textbf{b}
Control parameters of the network density of the tetra gel.
In the polymer network after completion of the chemical reaction, the molar mass of precursors $M$ corresponds to the molar mass between the crosslinks ($M/2$), and the polymer concentration $c$ represents the number density of crosslinks $n$ as $n=cN_{A}/M$.
Here, $N_{A}$ is the Avogadro constant.
}
\label{fig:1}
\end{figure}

In recent years, we overcame the difficulty of inhomogeneity by developing a tetra-arm poly(ethylene glycol) (PEG) hydrogel (tetra gel)~\cite{Sakai2008} with a homogeneous network structure~\cite{Matsunaga2009} (Fig.~\ref{fig:1}a).
In the tetra gel, we can independently and systematically control the structure of the polymer network, as shown in Fig.~\ref{fig:1}b.
Using tetra gels, we have studied the linear elasticity of polymer gels in the as-prepared state by various experimental techniques~\cite{Sakai2008, Akagi2013, Nishi2017, Yoshikawa2019}.\\

Until recently, we analyzed our experiments using the existing models of the classical rubber elasticity theories but observed inconsistencies with respect to the interpretations of experimental results as described in Sec.~\ref{sec:history}.
A very recent discovery revealed that polymer gels have ``negative energy elasticity"~\cite{Yoshikawa2021}, namely, a significant negative internal energy contribution to the shear modulus originating from the solvent.
This is not considered in the classical rubber elasticity theories, which assume that the elastic modulus is mainly determined from the entropy contribution.
Because the internal energy contribution is significant and negative, the (hidden) entropy contribution is large compared to the total modulus.
Our discovery challenges the conventional notion that elasticity of polymer gel can be understood by the classical rubber elasticity theories.\\

In this focus review, we describe how past experimental results on the linear elasticity of polymer gels can be successfully explained by the existence of negative energy elasticity.
The review is organized as follows.
First, we briefly review the experimental study on the linear elasticity of polymer gels.
Second, we present a current state-of-the-art unified formula for the linear elasticity of polymer gels with various network topologies and densities.
Third, using this formula, we re-examine the past experimental results.
Finally, we present the summary and outlook of these investigations.\\[47pt]

\begin{figure}[t!]
\centering
\includegraphics[width=0.89\linewidth]{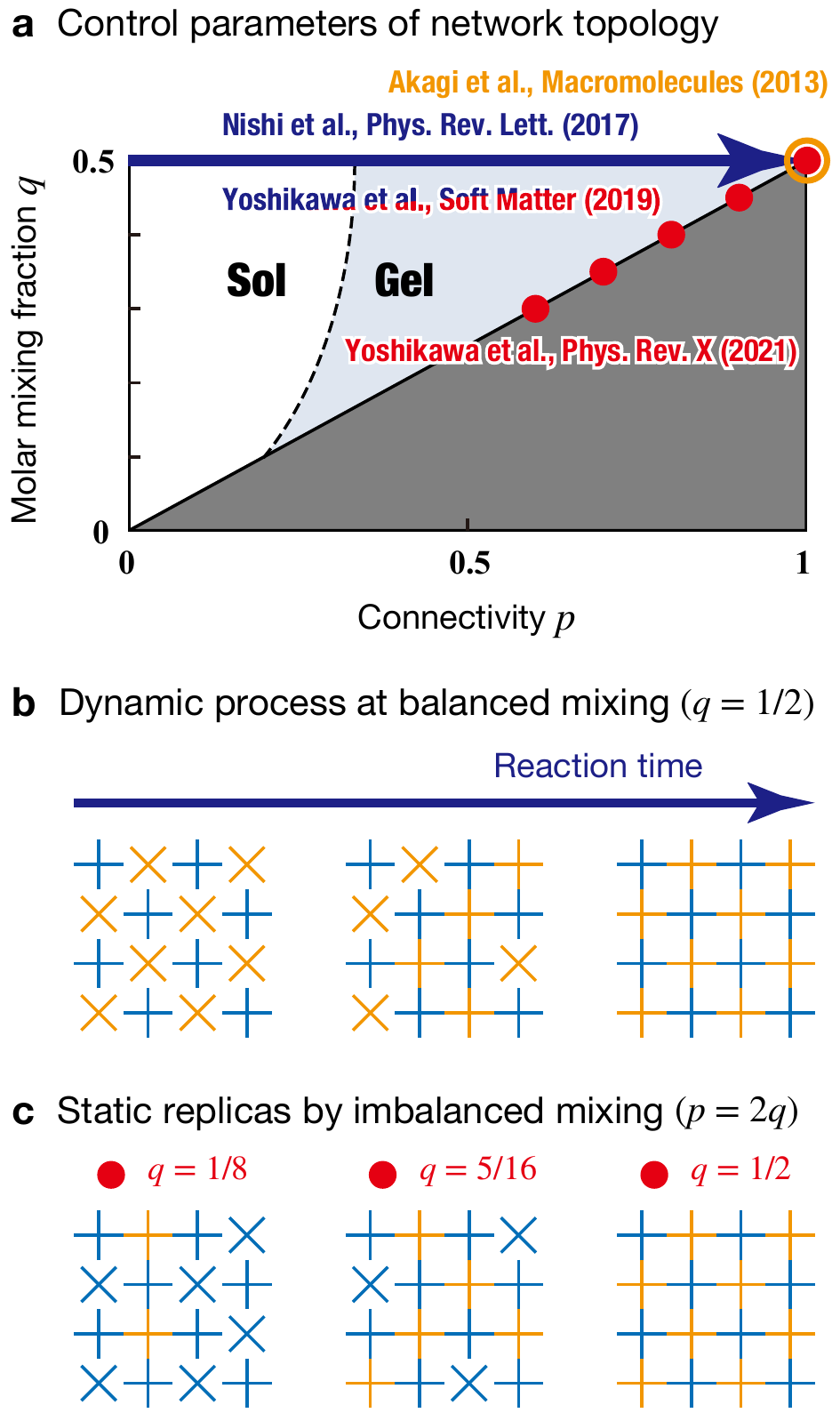}
\caption{
\textbf{a}
Control parameters of the network topology of the tetra gel.
The connectivity $p$ increases monotonically with time and (ideally) reaches $p=2q$ after completion of the reaction, where $q$ is the molar mixing fraction of the precursors of the minor group.\\
\textbf{b}
Dynamic process (DP) of gelation, where two precursors are mixed in a stoichiometrically balanced ratio ($q=1/2$).\\
\textbf{c}
Static replicas (SR) of DP, where two precursors are mixed in a stoichiometrically imbalanced (and balanced) ratio ($0\leq q\leq 1/2$).
Here, the connectivity $p$ after completion of the reaction is tuned as $p=2q$.
}
\label{fig:2}
\end{figure}

\begin{figure*}[t!]
\centering
\includegraphics[width=\linewidth]{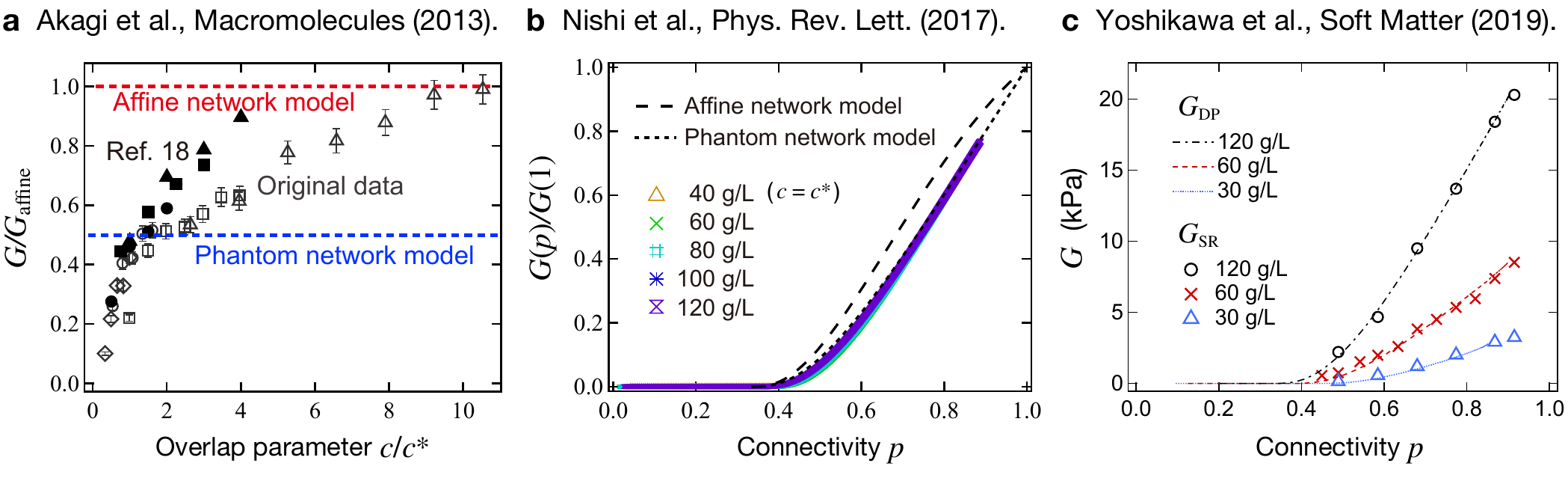}
\caption{
Representative experimental results before the discovery of negative energy elasticity.\\
\textbf{a} Normalized shear modulus ($G/G_\mathrm{affine}$) as a function of the overlap parameter ($c/c^*$).
The open gray symbols represent the data from the original paper~\cite{Akagi2013}, which are inaccurate due to the samples and measurement method.
The filled black symbols represent the data from Ref.~\cite{Yoshikawa2021} that are more accurate in terms of the samples and measurement method (see main text).
Rhombuses, circles, squares, and triangles represent $M=5,10,20$, and $40$ kg/mol, respectively.
The blue and red dashed lines show $G/G_\mathrm{affine}=0.5$ (the prediction of the phantom network model) and $1$ (the prediction of the affine network model), respectively.
The overlap parameter in the horizontal axis is converted from the polymer volume fraction $\phi/\phi^*$ (in the original paper~\cite{Akagi2013}) to concentration $c/c^*$.\\
\textbf{b} Normalized shear modulus ($G(p)/G(1)$) as a function of the connectivity of the polymer network ($p$).
The polymer concentrations are $c=40,60,80,100$, and $120$~g/L.
The molar mass of the precursors is $M=20$ kg/mol, and the corresponding overlap concentration is $40$ g/L.
We calculate the dashed lines from the affine and phantom network models with the Bethe approximation.
The data are taken from Ref.~\cite{Nishi2017}.\\
\textbf{c} Shear modulus ($G$) as a function of the connectivity of the polymer network ($p$) in the dynamic gelation process (DP) and the static replica (SR).
The polymer concentrations are $c=30$, $60$, and $120$~g/L, and the molar mass of the precursors is $M=20$ kg/mol.
The data are taken from Ref.~\cite{Yoshikawa2019}.
We note that the data in \textbf{b} and \textbf{c} were measured accurately in the same way as Ref.~\cite{Yoshikawa2021}.
}
\label{fig:3}
\end{figure*}

\section{Past Experimental Results of Linear Elasticity in Tetra Gels}
\label{sec:history}

In this section, we briefly review, in chronological order, our four experiments~\cite{Akagi2013,Nishi2017,Yoshikawa2019,Yoshikawa2021} that investigated the linear elasticity of polymer gels in the as-prepared state using tetra gels (Fig.~\ref{fig:1}a).
As shown in Fig.~\ref{fig:1}b, by tuning the molar mass $M$ and concentration $c$ of precursor solutions, we could independently and systematically control the network density, i.e., the molar mass between the crosslinks $M/2$ and the number density of crosslinks $n=cN_{A}/M$ in tetra gels.
Here, $N_{A}$ is the Avogadro constant, and $c$ is defined as the precursor weight divided by the solvent volume rather than by the solution volume (see Sec. S1 in Ref.~\cite{Yasuda2020}).
In addition, we could control the network topology by tuning the following two parameters:
(i) the connectivity $p$ ($0\leq p\leq1$), i.e., the fraction of the reacted terminal functional groups to all the terminal functional groups, and (ii) molar mixing fraction of minor precursors to all precursors $q$ as $[\mathrm{A}]:[\mathrm{B}]=q:1-q$ for $0\leq q\leq 1/2$.
Each of these experiments~\cite{Akagi2013,Nishi2017,Yoshikawa2019,Yoshikawa2021} involved different network topologies and is summarized in Fig.~\ref{fig:2}a.
Akagi et al.~\cite{Akagi2013} investigated networks with $p\simeq 1$ (after completion of the reaction) and $q=1/2$ (stoichiometrically balanced mixing), as shown by the orange circle in Fig.~\ref{fig:2}a.
Nishi et al.~\cite{Nishi2017} investigated networks with $q=1/2$, as shown by the blue arrow in Fig.~\ref{fig:2}a and Fig.~\ref{fig:2}b.
Yoshikawa et al.~\cite{Yoshikawa2019} compared networks with $q=1/2$ (blue arrow in \ref{fig:2}a and \ref{fig:2}b) and $p=2q$ (red filled circle in Fig.~\ref{fig:2}a and Fig.~\ref{fig:2}c).
Yoshikawa et al.~\cite{Yoshikawa2021} investigated networks with $q=1/2$ and $p=2q$ (red filled circle in Fig.~\ref{fig:2}a and Fig.~\ref{fig:2}c).
We describe the details of these studies in the following.

\newpage

The first two studies (Akagi et al.~\cite{Akagi2013} and Nishi et al.~\cite{Nishi2017}) investigated the applicability of the classical rubber theories to polymer-gel elasticity.
The representative models are the affine~\cite{Flory1953} and phantom~\cite{James1953} network models, which predict the shear modulus $G$ as 
\begin{equation}
G_\mathrm{affine}=\nu nk_{B}T
\label{eq:affine}
\end{equation}
and
\begin{equation}
G_\mathrm{phantom}=\xi nk_{B}T,
\label{eq:phantom}
\end{equation}
respectively.
Here, $n$, $k_{B}$, and $T$ are the number density of crosslinks, Boltzmann constant, and absolute temperature, respectively.
In Eqs.~(\ref{eq:affine}) and (\ref{eq:phantom}), $\xi\equiv \nu-\mu$ is the difference between the number per precursor of the elastically effective chains ($\nu$) and the crosslinks ($\mu$).
We cannot experimentally observe $\nu$ and $\xi$.
However, $p$ and $q$ can be observed and used to calculate the functions $\nu=\nu(p,q)$, $\mu=\mu(p,q)$, and $\xi=\xi(p,q)$ using the Bethe (i.e., tree) approximation~\cite{Macosko1976, Miller1976,Yoshikawa2019}.
The difference between these models (Eqs.~(\ref{eq:affine}) and (\ref{eq:phantom})) is the way they address the fluctuation of crosslinks.
The affine network model assumes that the crosslinks are fixed to the gel and that the deformation of a chain follows macroscopic deformation.
On the other hand, the phantom network model assumes that the crosslinks fluctuate and that the deformation of a chain is attenuated.

\newpage

Akagi et al.~\cite{Akagi2013} measured the $c$ and $M$ dependences of the shear modulus $G$ through stretching measurements of the network with $p\simeq 1$ and $q=1/2$.
(Strictly speaking, the connectivity $p$ of all the completely reacted gel samples was almost constant, $p\simeq 0.9$.)
Figure~\ref{fig:3}a demonstrates that all the data of the $c/c^{*}$ dependence of $G/G_\mathrm{affine}$ with different $M$ collapse onto a single master curve.
Here, $c^{*}$ is the overlap concentration of precursors obtained by viscosity measurement.
However, in the original paper~\cite{Akagi2013}, the measurement results (open gray symbols) were 
inaccurate for the following two reasons:
(i) a lower elastic modulus than expected was observed because the tetra gels were prepared using precursors with the terminal functional groups (amine and $N$-hydroxysuccinimide) that undergo hydrolysis over time;
(ii) the elastic modulus was measured by stretching measurement, which causes a large error.
To enable an accurate discussion, Fig.~\ref{fig:3}a also shows the accurately remeasured data from Ref.~\cite{Yoshikawa2021} (filled black symbols), overcoming the above two problems;
we (i) used tetra gels prepared using the precursors with terminal functional groups (maleimide and thiol) that do not cause hydrolysis and (ii) measured their elastic modulus by dynamic rheological measurement.
Here, the normalization factors $c^*=c^*(M)$ are different;
the original paper~\cite{Akagi2013} used $c^*=120, 75, 40, 15$ g/L for $M=5, 10, 20, 40$ kg/mol, respectively, whereas Ref.~\cite{Yoshikawa2021} used $c^*=60, 40, 30$ g/L for $M=10, 20, 40$ kg/mol, respectively.
We note that the following data (Figs.~\ref{fig:3}b and c below) were also accurately measured in the same way as Ref.~\cite{Yoshikawa2021}.\\

From our present understanding, Akagi et al. misinterpreted the results of Fig.~\ref{fig:3}a, i.e, a crossover from the phantom network model to the affine network model occurs in polymer gels.
In Fig.~\ref{fig:3}a, the horizontal line showing $G/G_\mathrm{affine} = 0.5$ can be regarded as the prediction of the phantom network model because
\begin{equation}
\frac{G_\mathrm{phantom}}{G_\mathrm{affine}}
=\frac{\xi(p,1/2)}{\nu(p,1/2)}
 \simeq 0.5,
\end{equation}
for a tetra-arm network at $p\simeq 1$.
For $c\simeq c^{*}$, 
$G$ agrees well with $G_\mathrm{phantom}$, and for $c<c^{*}$, the values of $G$ are smaller than those of $G_\mathrm{phantom}$.
This is probably due to an increase in ineffective connections for $c<c^{*}$~\cite{Yoshikawa2019}.
On the other hand, for $c>c^{*}$, $G/G_\mathrm{affine}$ increases to approach $1$ as $c$ increases.
Previously, it was considered for conventional polymer gels that an increase in $G/G_\mathrm{affine}$ with an increase in $c$ is due to the presence of trapped entanglements. 
However, the stress-elongation curve obeying the neo-Hookean model~\cite{Sakai2014} and the fracture energy obeying the Lake-Thomas model~\cite{Akagi2013f} strongly suggest that this is not the case. 
Therefore, Akagi et al. interpreted that the result in Fig.~\ref{fig:3}a indicates a crossover from the phantom network model to the affine network model with an increase in $c$.
However, this crossover is negated by the following $p$ dependence results.\\

\begin{figure*}[t!]
\centering
\includegraphics[width=0.64\linewidth]{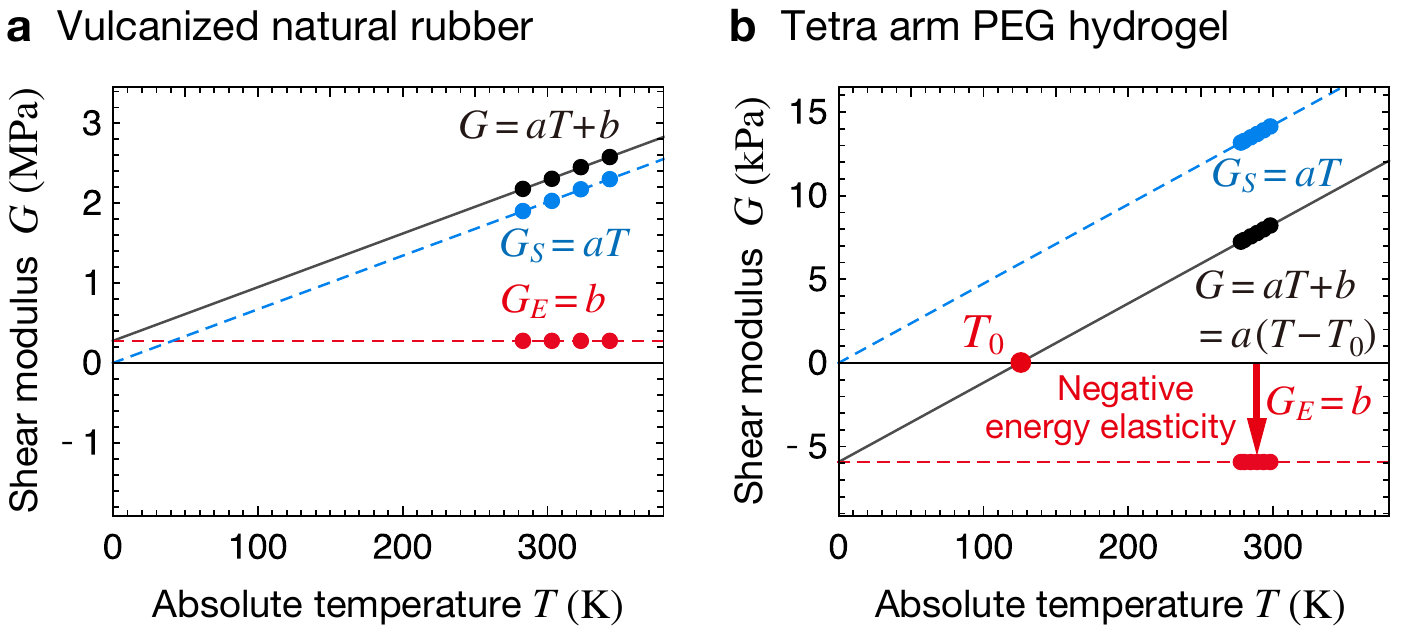}
\caption{
\textbf{a-b} Decomposition of entropy and energy contributions to shear modulus in (a) vulcanized natural rubber and (b) tetra gel.
We obtain the gray solid line from a least-squares fit to the temperature dependence of the shear modulus $G$ (black symbols).
According to Eq.~(\ref{eq:G-vantHoff}), we have the entropy contribution $G_{S}$ (blue dashed line) and the energy contribution $G_{E}$ (red dashed line), which corresponds to the intercept of the gray solid line.
The data are taken from Refs.~\cite{Anthony1942} and \cite{Yoshikawa2021} for \textbf{a} and \textbf{b}, respectively.
Notably, the shear modulus of vulcanized natural rubber is proportional to the absolute temperature ($G\simeq aT$), while that of the tetra gel is a linear function with a negative intercept [$G=a(T-T_0)$].
Here, the sample of tetra gel is synthesized by equal-weight mixing of the two kinds of precursors whose molar mass $M$ and concentration $c$ are $20$ kg$/$mol and $60$ g$/$L, respectively.
}
\label{fig:4}
\end{figure*}

\begin{figure*}[t!]
\centering
\includegraphics[width=0.95\linewidth]{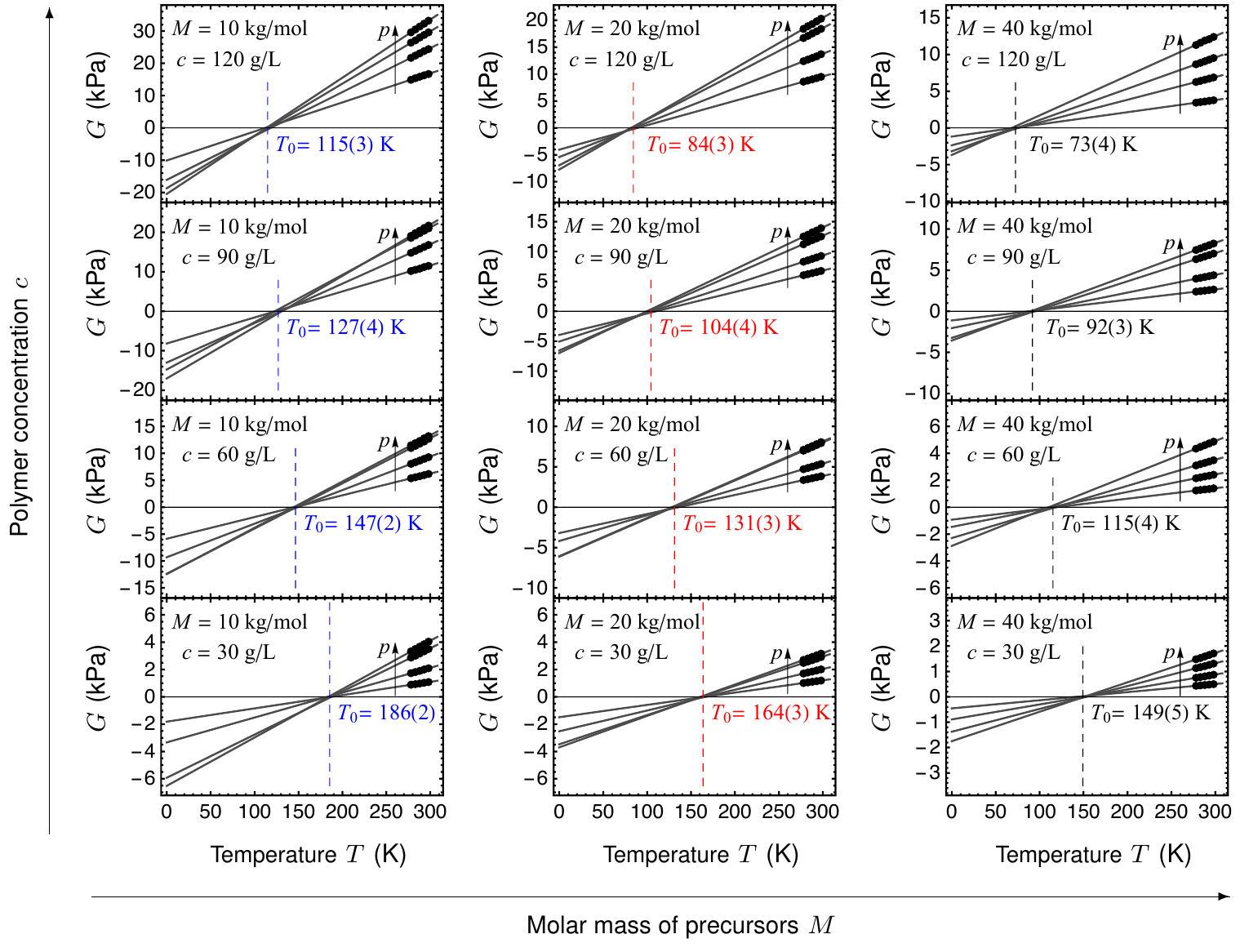}
\caption{
Experimental evidence for the existence of negative energy elasticity in a polymer gel.
All panels show the temperature ($T$) dependence of the shear modulus $G$.
We obtain each gray line from a least-square fit of each sample, which is characterized by the three parameters of the precursors:
the molar mass $M$, the concentration $c$ and the connectivity $p$.
All gray lines that have the same $M$ and $c$ pass through a vanishing temperature $T_0$ on the $T$ axis, which leads to Eq.~(\ref{eq:Gbya}).
The value of $T_0$ in each graph is the average of the four samples with different values of $p$, and the values in parentheses represent the standard deviation.
(Reprinted from Ref.~\cite{Yoshikawa2021}; CC BY 4.0.)
}
\label{fig:5}
\end{figure*}

\begin{figure*}[t!]
\centering
\includegraphics[width=0.95\linewidth]{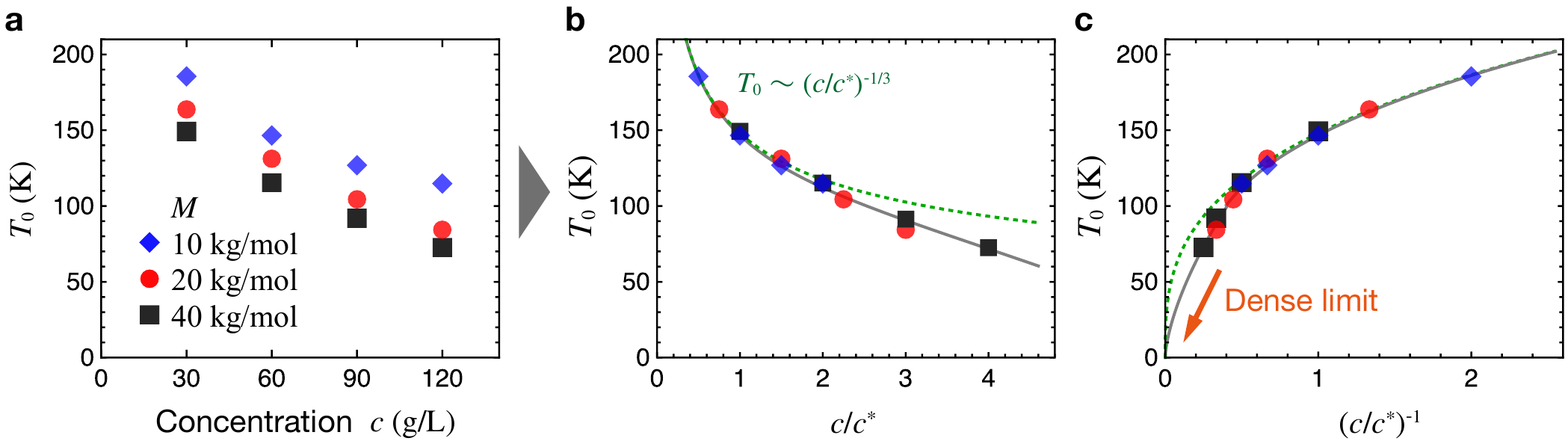}
\caption{
\textbf{a} The concentration ($c$) dependence of the vanishing temperature $T_0$ that governs the energy contribution of polymer-gel elasticity.
The blue diamonds, red circles, and black squares represent $M=10$, $20$, and $40$ kg/mol, respectively.
Each symbol represents the average of four samples taken from Fig.~\ref{fig:5} (i.e., the data are taken from Ref.~\cite{Yoshikawa2021}).
\textbf{b, c} The master curve of $T_0$ obtained by normalizing the concentration.
Here, we set $c^*= 60$, $40$, and $30$ g/L for $M=10$, $20$, and $40$ kg/mol, respectively.
The green dashed curve shows the scaling law $T_0\sim (c/c^*)^{-1/3}$ in the dilute regime ($c/c^*<1$).
As $(c/c^*)^{-1}\to 0$ (the dense limit), $T_0$ decreases, approaching nearly zero.
}
\label{fig:6}
\end{figure*}

Nishi et al.~\cite{Nishi2017} investigated $p$ dependence of $G$ in the range $c^{*}< c$ for a dynamic process (DP) in which a network is formed from two precursor solutions in a stoichiometric ratio ($q=1/2$), as shown by the blue arrow in Fig.~\ref{fig:2}b.
Just after mixing two precursor solutions, we measured the time ($t$) courses of (i) $G$ by rheological measurements and (ii) $p$ by ultraviolet-visible light spectroscopy.
Combining $G=G(t)$ and $p=p(t)$, we obtained $G(p)$. 
Figure~\ref{fig:3}b shows $G(p)/G(1)$ as a function of $p$, where $G(1)$ is the extrapolation of $G(p)$ at $p = 1$ based on the percolated network model~\cite{Nishi2015}.
Figure~\ref{fig:3}b demonstrates that all the data of the $p$ dependence of $G(p)/G(1)$ with different $c$ (in the range of $c^{*}< c$) collapse onto a single master curve, corresponding to the prediction of the phantom network model under the Bethe approximation, $G_\mathrm{phantom}(p)/G_\mathrm{phantom}(1)$.\\

Yoshikawa et al.~\cite{Yoshikawa2019} compared two methodologies to measure the connectivity ($p$) dependence of shear modulus $G$ in AB-type polymerization.
The first is to measure $G$ during the dynamic process (DP) of gelation in a stoichiometric ratio ($q=1/2$), as shown in Fig.~\ref{fig:2}b.
The second is to measure $G$ of samples whose $p$ after completion of the reaction is tuned by mixing two precursors in stoichiometrically imbalanced ($q<1/2$) and balanced ($q=1/2$) ratios.
Here, assuming the complete reaction of a minor group, we have $p=2q$.
This methodology can be regarded as a static replica (SR) of the DP, as shown in Fig.~\ref{fig:2}c.
In the former (DP), we obtain continuous $p$ dependence by monitoring the time evolution of the same sample, whereas in the latter (SR), we obtain discrete $p$ dependence by using different samples.
The advantage of the SR over the DP is that the SR can accurately measure various physical properties of samples with different $p$ over time because the system is static.
Figure~\ref{fig:3}c demonstrates that $G=G(p)$ in the two methodologies (DP and SR) agrees well in high $p$ (i.e., $p > 0.75$).
[In low $p$, close to the gelation point, the differences in the structural parameters become more pronounced, reflecting the differences in the topology of the DP (e.g., $\xi(p,1/2)$) and SR (e.g., $\xi(p,p/2)$).
See Fig.~4 in Ref.~\cite{Yoshikawa2019}.]
Note that these behaviors of the $p$-dependence of $G$ in the DP and SR are well reproduced by the phantom network model under the Bethe approximation.\\

The above series of studies~\cite{Akagi2013,Nishi2017,Yoshikawa2019} are based on the longstanding basic assumption that the polymer-gel elasticity is mainly determined by the entropy contribution.
Under this assumption, polymer-gel elasticity has been evaluated with the classical rubber elasticity theories~\cite{Flory1953,James1953,Flory1977} that predict that the shear modulus is proportional to the absolute temperature ($G\simeq aT$), such as Eqs.~(\ref{eq:affine}) and (\ref{eq:phantom}).
As shown in Fig.~\ref{fig:4}a, many experimental studies on natural and synthetic rubbers~\cite{Meyer1935,Anthony1942,Mark1965,Mark1976} have confirmed that $G\simeq aT$, which means that $G$ is mainly determined by the entropy contribution.
However, in the case of a polymer gel, the results analyzed using such an assumption ($G\simeq aT$) were found to be inconsistent, even for measurements of $G$ at certain temperatures (room temperature).
For example, the results shown in Figs.~\ref{fig:3}b and c seem to be inconsistent with the result of Fig.~\ref{fig:3}a; 
while Fig.~\ref{fig:3}a shows the crossover between the phantom and affine network models depending on $c$, Figs.~\ref{fig:3}b and c are consistent with the phantom network model not depending on $c$.
We cannot reconcile this inconsistency as long as we assume $G\simeq aT$.\\

Yoshikawa et al.~\cite{Yoshikawa2021} examined whether the premise of $G\simeq aT$
 [e.g., Eqs.~(\ref{eq:affine}) and (\ref{eq:phantom})] holds for polymer gels by measuring the temperature ($T$) dependence of the shear modulus $G$.
Taking advantage of SR, we prepared various gel samples with different network densities (various $M$ and $c$) and network topologies (various $p$).
As shown in Fig.~\ref{fig:4}b, we found that $G=aT+b$ with a significant negative value of $b$, contrary to the premise of the classical rubber elasticity theories.
As we explain in the next section, the first and second terms ($aT$ and $b$) correspond to entropy and internal energy contributions to $G$, respectively.
Thus, the negative value of $b$ is interpreted as ``negative energy elasticity''.
In Ref.~\cite{Yoshikawa2021}, we confirm the above conclusions with more than 50 different network topologies and densities (Fig.~\ref{fig:5}).
In the next section, we explain the negative energy elasticity based on thermodynamics and provide a self-contained description of the unified formula, which can explain all the experimental results of the linear elasticity of PEG hydrogels with various network topologies and densities~\cite{Akagi2013,Nishi2017,Yoshikawa2019,Yoshikawa2021}.

\section{Unified Formula for Linear Elasticity of Polymer Gels}

This section provides a self-contained summary of the state-of-the-art understanding of the elasticity of isotropic, incompressible gels in the as-prepared state, as obtained using tetra gels with a homogenous network.
In general, for such homogeneous isotropic linear elastic materials, the elastic properties are uniquely determined by the shear modulus $G$, i.e., any of the other elastic moduli can be calculated from $G$.
For example, Young's modulus $E$ can be calculated as $E=3G$, and the bulk modulus $K$ is considered infinite ($K \gg G$).
Therefore, the elasticity of polymer gels in the as-prepared state is entirely determined by $G$.\\

We consider the thermodynamics of the deformation of polymer gels in the as-prepared state.
The derivative of the Helmholtz free energy $F$ of an elastic body with an applied shear strain $\gamma$ is given by
$dF = -SdT -PdV + V\sigma d\gamma$
at temperature $T$ and external pressure $P$~\cite{LandauLifshitz,Flory1953}.
Here, $S$, $V$, and $\sigma$ are the entropy, volume, and shear stress, respectively.
Polymer gels are considered to be incompressible, i.e., the relative volume change $\Delta V/V$ is negligible because the bulk modulus (on the order of GPa) is significantly larger than the shear modulus (on the order of kPa).
(We present detailed analysis of $\Delta V/V$ in Appendices~B and C in Ref.~\cite{Yoshikawa2021}.)
Thus, we have
\begin{equation}
df = -sdT + \sigma d\gamma,
\label{eq:dF}
\end{equation}
where $f\equiv F/V$ and $s\equiv S/V$ are the Helmholtz free energy and entropy densities, respectively.\\

In polymer physics, $f$ of a polymer gel is often written in the form of two separate contributions as~\cite{Flory1953}
\begin{equation}
f\left(T,\gamma\right)
=f_{\mathrm{mix}}\left(T\right)
+f_{\mathrm{el}}\left(T,\gamma\right),
\label{eq:Ftot}
\end{equation}
where 
$f_{\mathrm{mix}}\left(T\right)\equiv f\left(T,0\right)$ and 
$f_{\mathrm{el}}\left(T,\gamma\right) \equiv f\left(T,\gamma\right)-f_{\mathrm{mix}}\left(T\right)$
are the mixing and elastic free energy densities, respectively.
Here, $f_{\mathrm{mix}}$ is independent of the applied shear strain $\gamma$ because the volume $V$ does not change with deformation.
We emphasize that Eq.~(\ref{eq:Ftot}) does not provide any new information in the as-prepared state; it merely defines $f_{\mathrm{mix}}$ and $f_{\mathrm{el}}$.
Equation~(\ref{eq:dF}) gives the shear stress as $\sigma (T,\gamma) = \partial f (T,\gamma)/\partial \gamma$ in an isothermal process.
Thus, the shear modulus 
($G(T)\equiv \lim_{\gamma\to 0} \partial \sigma (T,\gamma)/\partial \gamma$)
is related to the free energy as
\begin{equation}
G\left(T\right)
\equiv 
\lim_{\gamma\to 0} \frac{\partial^{2} f}{\partial \gamma^{2}} (T,\gamma)
=
\lim_{\gamma\to 0}
\frac{\partial^{2} f_{\mathrm{el}}}{\partial \gamma^{2}} (T,\gamma).
\label{eq:G-Fel}
\end{equation}
Equation~(\ref{eq:G-Fel}) indicates that $f_{\mathrm{mix}}$ does not contribute to the shear modulus $G$.\\

Recently, we obtained a unified expression of the shear modulus of tetra gels as a function of the microscopic structure of the polymer network as~\cite{Yoshikawa2021}
\begin{equation}
G(T;c,M,p,q)=a(c,M,p,q) \left[ T-T_0\left(\frac{c}{c^*(M)}\right)\right],
\label{eq:Gbya}
\end{equation}
where $c$, $M$, $p$, and $q$ are the polymer concentration, molar mass of precursors, connectivity, and molar mixing fraction, respectively.
(The definitions of $p$ and $q$ are given in the previous section.)
Figures~\ref{fig:5}, \ref{fig:6}a, and \ref{fig:6}b experimentally validate Eq.~(\ref{eq:Gbya}) as follows.
First, Fig.~\ref{fig:5} shows that $G$ is a nearly linear function of $T$ [i.e., $G=aT+b=a(T-T_0)$, where $T_{0}=-b/a$] over the measured range ($278\, \mathrm{K}\leq T \leq 298\, \mathrm{K}$).
Second, Figs.~\ref{fig:5} and \ref{fig:6}a show that $T_0$ does not depend on the network topology ($p$ and $q$) but depends on the network density ($c$ and $M$).
Finally, Fig.~\ref{fig:6}b demonstrates that the dependence of $c$ and $M$ on $T_{0}$ is governed by $T_{0}=T_{0}(c/c^*(M))$.
Here, $c^*(M)$ is the normalization factor chosen to construct the master curve.
We note that $c^*(M)$ is in close agreement with the overlap concentration of the precursors $c_{\mathrm{vis}}^*(M)$ obtained by the viscosity measurement~\cite{Akagi2013}.\\

The first and second terms in Eq.~(\ref{eq:Gbya}) correspond to the entropy and energy elasticity, respectively.
The Helmholtz free energy density satisfies $f=e-Ts$, where $e$ is the internal energy density and $s$ is the entropy density.
Thus, on the basis of Eq.~(\ref{eq:G-Fel}), we define the energy contribution $G_E$ and the entropy contribution $G_S$ to the shear modulus $G=G_E+G_S$ as
$G_{E} \equiv\lim_{\gamma\to 0}\left(\partial^2 e/\partial \gamma^2\right)_{T,V}$
and
$G_{S} \equiv  -\lim_{\gamma\to 0}T\left(\partial^2 s/\partial \gamma^2\right)_{T,V}$, respectively.
Here, $G_{S}$ and $G_{E}$ are defined under a constant-volume condition.
According to the Maxwell relation $\left(\partial s/\partial \gamma\right)_{T,V} = -\left(\partial \sigma/\partial T\right)_{\gamma,V}$, we have
\begin{equation}
G_{S} (T)=T\frac{dG}{dT} (T),
\label{eq:G-vantHoff}
\end{equation}
which enables us to determine the entropy and energy contributions from the temperature ($T$) dependence of the shear modulus $G$ under a constant-volume condition~\cite{Fermi1937,Flory1953, Rubinstein2003}.
Substituting Eq.~(\ref{eq:Gbya}) into Eq.~(\ref{eq:G-vantHoff}), we have $G_{S}=a T$ and $G_{E}=G-G_{S}=-aT_{0}$.\\

All measured samples shown in Fig.~\ref{fig:5} have a significant negative $G_{E}$, which indicates that the undeformed state is unstable in terms of the internal energy.
Because the (total) shear moduli of stable materials are generally bound to be positive ($G>0$), $G_{S}=aT$ must be larger than $\left|G_{E}\right|=aT_{0}$.\\

We discuss Eq.~(\ref{eq:Gbya}) in the dilute, semidilute, and dense regimes, respectively.
In the dilute regime ($c/c^*(M)<1$), Fig.~\ref{fig:6}b demonstrates the scaling law $T_0\sim \left(c/c^*(M)\right)^{-1/3}$, which would be key to further understanding the microscopic origin of negative energy elasticity in the future.
We present further discussion in Ref.~\cite{Yoshikawa2021}.\\

In the semidilute regime ($1\lesssim c/c^*(M)\lesssim 4$) and sufficiently high $p$ (i.e., $p > 0.75$), our experiment~\cite{Yoshikawa2019,Yoshikawa2021} shows that the entropy contribution $G_S$ of the tetra gel is phenomenologically well reproduced by a constant multiple of the phantom network model [Eq.~(\ref{eq:phantom})] as $G_S \simeq 2.4G_\mathrm{phantom}\equiv 2.4 \xi nk_{B}T$,
where $n=n(c,M)=cN_{A}/M$ is the number density of the tetra-arm precursors.
This implies that the prefactor $a(c,M,p,q)$ is approximately separable into the product of the network density contribution (with control parameters $c$ and $M$, Fig.~\ref{fig:1}b) and the network topology contribution (with control parameters $p$ and $q$, Fig.~\ref{fig:2}a) as
\begin{equation}
a(c,M,p,q) \simeq 2.4 k_{B} n(c,M) \xi(p,q).
\label{eq:phantom-like}
\end{equation}
We note that Eq.~(\ref{eq:phantom-like}) is valid in the semidilute regime ($1\lesssim c/c^*(M)\lesssim 4$) and sufficiently high $p$.
In fact, $a(c,M,p,q)$ is a complex function in the dilute regime ($c/c^*(M)<1$) or low $p$ [see Fig.~5(d) in Ref.~\cite{Yoshikawa2021}].\\

In the dense regime, Fig.~\ref{fig:6}c indicates that $T_0$ decreases, approaching nearly zero as $(c/c^{*})^{-1} \to 0$, which means that the solvent is removed.
This result is consistent with experimental results on vulcanized natural rubber and synthetic rubbers without solvent; the absolute value of $G_{E}$ is much smaller than the value of $G_{S}$~\cite{Meyer1935,Mark1965,Mark1976}. 
In other words, this result suggests that the negative energy elasticity in the polymer gels originates from the solvent.

\section{Re-examination of Past Experimental Results by the Unified Formula}

Based on the unified formula in Eq.~(\ref{eq:Gbya}) together with the phenomenological expression of the prefactor in Eq.~(\ref{eq:phantom-like}), we re-examine past experimental results on polymer-gel elasticity~\cite{Akagi2013,Nishi2017,Yoshikawa2019}.
First, we consider Akagi et al.~\cite{Akagi2013}, which investigated the network with $p\simeq 1$ and $q=1/2$, as shown by the orange circle in Fig.~\ref{fig:2}a.
Substituting Eq.~(\ref{eq:phantom-like}) into Eq.~(\ref{eq:Gbya}) and using Eq.~(\ref{eq:affine}), we have
\begin{equation}
\begin{split}
\frac{G\left(T;c,M,1,\frac{1}{2}\right)}{G_\mathrm{affine}\left(T;c,M,1,\frac{1}{2}\right)}
& \simeq \frac{2.4\xi\left(1,\frac{1}{2}\right)
\left[ T-T_0\left(\frac{c}{c^*(M)}\right)\right]}{\nu\left(1,\frac{1}{2}\right)T}\\
& \simeq 1.2 \left[1-\frac{T_0\left(\frac{c}{c^*(M)}\right)}{T}\right].
\label{eq:PRX-Akagi}
\end{split}
\end{equation}
Here, we use $\xi\left(1,1/2\right)=1$ and $\nu\left(1,1/2\right)=2$.
Equation~(\ref{eq:PRX-Akagi}) shows that $G/G_\mathrm{affine}$ depends only on $c/c^*$ under isothermal conditions (i.e., $T$ is constant), which elucidates Fig.~\ref{fig:3}a.
We emphasize that the ``crossover'' in Fig.~\ref{fig:3}a originates not from the phantom-affine crossover but from the concentration dependence of the negative energy elasticity (i.e., the dependence of $T_0$ on $c/c^*$).

\newpage

Second, we consider the findings of Nishi et al.~\cite{Nishi2017} on the network with $q=1/2$, as shown by the blue arrow in Fig.~\ref{fig:2}a.
As shown in Fig.~\ref{fig:3}b, Nishi et al.~\cite{Nishi2017} considered $G(p)/G(1)$, i.e., the shear modulus $G(p)$ normalized by the modulus of the network with $p= 1$ and $q=1/2$.
To generalize this, we consider $G(p,q)/G(1,1/2)$, i.e., the shear modulus $G(p,q)$ with general $q$ normalized by the modulus of the network with $p=1$ and $q=1/2$.
Substituting Eq.~(\ref{eq:phantom-like}) into Eq.~(\ref{eq:Gbya}), we have
\begin{equation}
\frac{G(T;c,M,p,q)}{G(T;c,M,1,1/2)}
=\frac{a(c,M,p,q)}{a(c,M,1,1/2)}
\simeq \xi(p,q).
\label{eq:PRX-Nishi}
\end{equation}
Equation~(\ref{eq:PRX-Nishi}) (with $q=1/2$) fully explains the result of Fig.~\ref{fig:3}b.
The ratio $G(T;c,M,p,q)/G(T;c,M,1,1/2)$ does not depend on $T$, $c$, and $M$, and is explained by $\xi(p,q)$, corresponding to the prediction of the phantom network model under the Bethe approximation.
In the previous section, we mentioned that the results of Akagi et al.~\cite{Akagi2013} and Nishi et al.~\cite{Nishi2017} seem to be inconsistent with each other.
However, the unified formula [Eq.~(\ref{eq:Gbya}) with Eq.~(\ref{eq:phantom-like})]
can explain both in a consistent manner.\\

Third, we consider Yoshikawa et al.~\cite{Yoshikawa2019}, which compared the networks with $q=1/2$ (blue arrow in Fig.~\ref{fig:2}a) and $p=2q$ (red filled circle in Fig.~\ref{fig:2}a).
The former is the DP, as shown in Fig.~\ref{fig:2}b, and the latter is the SR, as shown in Fig.~\ref{fig:2}c.
[In Ref.~\cite{Yoshikawa2019}, we referred to the SR as imbalanced mixing (IM).]
From the unified formula [Eq.~(\ref{eq:Gbya}) with Eq.~(\ref{eq:phantom-like})], we derive that the connectivity ($p$) dependence of the shear modulus of the DP ($G_\mathrm{DP}(p)$) and that of the SR ($G_\mathrm{SR}(p)$) agree well in high $p$ (i.e., $p > 0.75$), as shown in Fig.~\ref{fig:3}c.
Because $G_\mathrm{DP}(p)\equiv G(T;c,M,p,1/2)$ and $G_\mathrm{SR}(p)\equiv G(T;c,M,p,p/2)$, we obtain
\begin{equation}
\frac{G_\mathrm{SR}(p)}{G_\mathrm{DP}(p)}
\simeq \frac{\xi_\mathrm{SR}(p)}{\xi_\mathrm{DP}(p)},
\label{eq:PRX-Softmatter}
\end{equation}
where
\begin{gather}
\xi_\mathrm{DP}(p)\equiv \xi(p,1/2)=2p-1+O\left(\left(1-p \right)^{2}\right)
\label{eq:PRX-Softmatter-DP}\\
\xi_\mathrm{SR}(p)\equiv \xi(p,p/2)=2p-1+O\left(\left(1-p \right)^{2}\right).
\label{eq:PRX-Softmatter-SR}
\end{gather}
Here, we use the Bethe approximation~\cite{Yoshikawa2019}.
From Eqs.~(\ref{eq:PRX-Softmatter}), (\ref{eq:PRX-Softmatter-DP}), and (\ref{eq:PRX-Softmatter-SR}), we have $G_\mathrm{DP}(p) \simeq G_\mathrm{SR}(p)$ for $p\simeq 1$.
The essence of this result relies on the fact that the prefactor $a(c,M,p,q)$ is separable as in Eq.~(\ref{eq:phantom-like}), which leads to
\begin{equation}
\frac{G(T;c,M,p,q)}{\xi(p,q)}
\simeq 2.4 k_{B} n(c,M) \left[ T-T_0\left(\frac{c}{c^*(M)}\right)\right].
\end{equation}
Thus, $G/\xi$ is independent of network topology ($p$ and $q$).
In fact, Yoshikawa et al.~\cite{Yoshikawa2019} experimentally demonstrated that
$G_\mathrm{DP}(p)/\xi_\mathrm{DP}(p)=G_\mathrm{SR}(p)/\xi_\mathrm{SR}(p)$ holds and that $G_\mathrm{DP}(p)/\xi_\mathrm{DP}(p)$ (and $G_\mathrm{SR}(p)/\xi_\mathrm{SR}(p)$) does not depend on $p$ in the semidilute ($1\lesssim c/c^*(M)\lesssim 4$) and sufficiently high $p$.

\section{Summary and Future Prospects}

In this article, we have described how past experimental results~\cite{Akagi2013,Nishi2017,Yoshikawa2019} on the linear elasticity of polymer gels can be successfully explained by considering negative energy elasticity coexisting with entropy elasticity~\cite{Yoshikawa2021}.
First, we have reviewed the experimental research~\cite{Akagi2013,Nishi2017,Yoshikawa2019,Yoshikawa2021} on the linear elasticity of polymer gel in the as-prepared state using tetra gels.
Each of these experiments involves a different network topology (Fig.~\ref{fig:2}a).
Figure~\ref{fig:3} shows representative experimental results.
Second, we have provided the unified formula in Eq.~(\ref{eq:Gbya}) for the linear elasticity of polymer gels, which has been revealed in a series of studies~\cite{Akagi2013,Nishi2017,Yoshikawa2019,Yoshikawa2021}.
In addition, in the semidilute ($1\lesssim c/c^*(M)\lesssim 4$) and sufficiently high $p$, the prefactor $a(c,M,p,q)$ is separable as in Eq.~(\ref{eq:phantom-like}).
Finally, using this unified formula~(\ref{eq:Gbya}) together with the phenomenological expression of the prefactor in Eq.~(\ref{eq:phantom-like}), we have explained the past experimental results~\cite{Akagi2013,Nishi2017,Yoshikawa2019} and have reconciled the past results that seem to be inconsistent with each other~\cite{Akagi2013,Nishi2017}.\\ \\

The discovery of negative energy elasticity~\cite{Yoshikawa2021} is one of the most significant recent advances in the field of linear elasticity of polymer gels.
This negative energy elasticity, which vanishes when the solvent is removed, is the critical factor that differentiates gels from rubbers.
The models of classical rubber elasticity theories (such as the affine, phantom, and junction affine network models) are inapplicable to polymer gels; the relative change in shear modulus due to changes in temperature is several times greater than predicted by these models.
Thus, the negative energy elasticity is of great practical importance because polymer gels are used at various temperatures.\\ \\

The elucidation of the governing law of the shear modulus is expected to improve our understanding of the static and dynamic properties of swelling of polymer gels in solvents.
This is because the swelling pressure ($\Pi_{\mathrm{tot}}$) is determined by the shear modulus and osmotic pressure as $\Pi_{\mathrm{tot}}=\Pi_{\mathrm{mix}}+\Pi_{\mathrm{el}}$, where $\Pi_{\mathrm{mix}}$ and $\Pi_{\mathrm{el}}$ are the solvent-polymer mixing ($\Pi_{\mathrm{mix}}$) and elastic ($\Pi_{\mathrm{el}}$) contributions, respectively.
As an example of the static properties of swelling, we recently discovered the governing law of osmotic pressure throughout the gelation process involving both the sol and gel states~\cite{Yasuda2020}. 
Here, the osmotic pressure was obtained by controlling the swelling pressure of gels with an external solution and subtracting the elastic contribution ($\Pi_{\mathrm{el}}$).
As an example of the dynamic properties of swelling, we demonstrated a significant dependence of the elastic modulus on the collective diffusion coefficient of the polymer networks~\cite{Fujiyabu2019,Kim2020}.
Accordingly, elasticity is the basis of both static and dynamic properties of polymer gels, prompting us to revisit past previous studies.

\newpage

A complete understanding of the linear elasticity of polymer gels is still elusive, and it is important to address the following three points.
First, the microscopic origin of the negative energy elasticity needs to be clarified.
In Ref.~\cite{Yoshikawa2021}, we inferred that the origin is the interaction between the polymer and the solvent.
However, because we performed only macroscopic measurements~\cite{Akagi2013,Nishi2017,Yoshikawa2019,Yoshikawa2021}, we require further investigations to clarify the microscopic origin.
For example, molecular-scale experiments (such as light scattering and single-chain experiments~\cite{Nakajima2006,Liang2017}) and numerical simulations (such as molecular dynamics simulations) will reveal the microscopic origin.
A future theory that explains the origin of negative energy elasticity from the microscopic point of view should predict the vanishing temperature $T_{0}$ from the network structure.
To construct such a theory, the scaling law $T_0\sim \left(c/c^*(M)\right)^{-1/3}$ in the dilute regime ($c/c^*(M)<1$) in Fig.~\ref{fig:6}b would play a key role.\\

Second, in connection with the prefactor $a(c,M,p,q)$, we need to revisit the classical rubber elasticity theories, which are used to calculate the entropy elasticity of various network models.
The entropy elasticity ($G_{S}= G-G_{E}$), rather than the shear modulus itself ($G$), may be explainded by some classical rubber elasticity theory.
However, according to our experiments~\cite{Yoshikawa2021}, $G_{S}$ is $2.4$ times that of the phantom network model, as in Eq.~(\ref{eq:phantom-like}).
The universality and meaning of $2.4$ is an important open question.
For example, a recently proposed theory~\cite{Zhong2016} that predicts $G_{S}$ from the number of topological loop defects does not appear to explain our result because it predicts a lower $G_{S}$ than the phantom network model.

Third, it is important to investigate whether our findings extend to other polymer gels as well:
(i) homogeneous gels with other polymer-solvent systems, such as PEG-acetonitrile~\cite{Li2019}, poly(acrylic acid) (PAA)-water~\cite{Oshima2014}, poly($N$-isopropylacrylamide) (PNIPA)-methanol~\cite{Okaya2020}, and poly($n$-butyl acrylate) (PBA)-$N$,$N$-dimethylformamide (DMF)~\cite{Huang2020,Nakagawa2021} systems;
(ii) gels with other network structures, such as near-critical gels~\cite{Aoyama2021}
 and topological gels~\cite{Ito2007}; and
(iii) gels synthesized by other types of polymerization, such as radical polymerization.
We emphasize that $G_{S}$ and $G_{E}$ are defined (and Eq.~(\ref{eq:G-vantHoff}) holds) under constant-volume conditions.
Thus, it is necessary to confirm that the effect of volume changes due to temperature changes is negligible.
It is difficult for a small group to experiment on all of these various polymer gels, and it is essential to investigate the linear elasticity by various groups.

\begin{acknowledgments}
This work was supported by the Japan Society for the Promotion of Science (JSPS) through Grants-in-Aid for
Early Career Scientists grant number 19K14672 to N.S,
JSPS Fellows grant number 21J13478 to Y.Y,
Scientific Research (B) grant number 18H02027 to T.S,
Scientific Research (A) grant number 21H04688 to T.S,
and Transformative Research Areas grant number 20H05733 to T.S.
This work was also supported by the Japan Science and Technology Agency (JST) CREST grant number JPMJCR1992 to T.S.\\
\end{acknowledgments}


\end{document}